# Laval Turbonozzle Dynamics

S. L. Arsenjev, I. B. Lozovitski[1], Y. P. Sirik

*Physical-Technical Group*
*Dobrolubova Street, 2, 29, Pavlograd, Dnepropetrovsk region, 51400 Ukraine*

The results of computing experiments are presented for the steam and gas stream motion in converging-diverging flow element called Laval nozzle and applied in steam and gas turbines. In beginning the experiments had carried out for Laval nozzle tested by Stodola and then for this turbonozzle various modifications. Quantitative evaluation of the state and motion parameters of the viscous compressible fluid and the metering characteristics of the flow elements are given as well as physically adequate interpretation of the obtained results is presented at the first time. The computing experiments are carried out by means of VeriGas program-modeler composed on the base of a new approach to the classical physics development in field of fluid motion.

**Introduction**

A property of the converging-diverging flow element to increase the water flow is well known at least since Roman Empire. Approximately in beginning of the 21st century such flow elements became known as G.Venturi nozzle. Creation and development thermal engines – in beginning the steam and then internal combustion – favored that in 1848 Van Raten has established a possibility to obtain the gas stream supersonic velocity (Engl. Patent № 11800 ) by means of the Venturi nozzle. After roughly 40 years G. Laval has understood that by means of such nozzle it is possible more fully to utilize the steam stream energy for creation of motive power in comparison with the simple converging nozzle and he has included the supersonic nozzle into his steam turbine construction (Engl. Patent № 7143 ).

Such single-stage active turbines are applied in the various auxiliary units without essential changes now. Approximately at the same time C. Parsons has created the multi-stage steam turbine in which the steam is expanded both in supersonic nozzle and in the turbine wheels. Such steam turbine is the main kind of the driving gear for the electric power plants now. Perfection of both type of steam turbines as well as gas turbines created in the 20th century and also contained supersonic nozzles is now. And as in former times the perfection of turboplant construction is carried out by means of industrial experiment of models and pilot plants. This way of perfection is very expensive, long and one does not vouch for success. To create a possibility for physically adequate designed-theoretical reproduction of the state and motion parameters of the viscous compressible fluid stream and for determination of the supersonic nozzle metering characteristics on the design stage and in this way to use the computing experiment instead industrial is one of typical problems of turbine construction now.

---

[1] Phone: +44 7891943934 (Engl)
    +38 05632 40596 (Rus.)
E-mail: loz@inbox.ru

## General formulation

General formulation to the problem solution on determining of the viscous compressible fluid stream parameters and the flow element metering characteristics with taking into account of a friction and the cross-section area change is stated in author's article [1]. One of features of this problem is that the turbonozzle is a delivery nozzle i. e. an active nozzle. Thus its energy characteristics are a stream force determined by a product of the running quantity of velocity head and the corresponding cross-section area of stream as well as the specific impulse determined by a ratio of the stream force to its weight flow. Other feature is that the stream motion can be under action of the increased pressure before inlet of nozzle and under action of the decreased backpressure behind outlet of nozzle. Thus the nozzle metering characteristics is to determine for both mentioned cases.

Besides that, the question on influence of the nozzle diverging part profile and the wall micro-profile of the part onto the stream parameters and the stream nozzle energy leaves open. Now, it is known at least three types of the turbonozzle diverging part profile: straight (conical) by Laval, curvilinear concave by T. Stanton and curvilinear convex by Frankl [2].It is also known on various modifications of the listed profiles. Thus the qualitative evaluation of influence of geometrical features of the nozzle diverging part profile and its size onto the stream parameters and the stream nozzle energy is to give. Authors accept the initial profile of the nozzle diverging part and the stream pressure maximum drop close by A. Stodola's tests [3, 4].

## Solution

The problem solution has carried out by means of VeriGas program-modeler today's version which makes a computation in one-dimensional statement of problem. This approach can raise doubts in rightness of computation results, but so essential on the face of it geometrical simplification of the problem statement is compensated by physical adequacy of the problem solution algorithm. Authors by own experience of successful solution of a number of technical problems are convinced of the high effectiveness of the simplified approaches to its solutions for an engineering practice on condition that the understanding of the flow phenomenon is physical adequate. At the same time, authors develop the consequent versions of VeriGas program-modeler in axisymmetric statement and then they envisage 3-D statement. These subsequent versions will allow enriching the solution by new details, but they will not change the developed physical basics.

Results of solution of the problem in the given article are presented in the kind of graphs. The change of the state and motion parameters of steam stream along the length of the different profile and length nozzles, beginning with Stodola's profile, as well as a change of the velocity head, the velocity head force and specific impulse are adduced on Fig. 1 – 6. The weight flow and the stream specific impulse are given in bottom of the mentioned graphs. The metering characteristics of Stodola's turbonozzle (Fig. 1) and its inlet part are separately presented on Fig. 7, 8. The metering characteristics are given separately for a case of increase of pressure before inlet of nozzle when backpressure remains invariable – left graphs and for a case of decrease of backpressure when pressure before inlet of nozzle remains invariable – right graphs.

## Discussion of results

Feature of the computation results of the state and motion parameters of a steam stream in the nozzle diverged part, presented on Fig. 1, is in uncommon character of the temperature and velocity curves. The feature is a consequence that the computation of the viscous compressible fluid stream parameters were carried out with tacking into account of a friction and one can be explained in the following way. The steam stream moves in the nozzle diverging part at supersonic velocity. Friction of supersonic stream against the nozzle wall is accompanied by heating of the stream analogically to the subsonic stream friction. But in contrast to the latter a heating of supersonic stream leads to it's braking. Thus the stream velocity is decreased, sonic velocity is increased in it and Mach number is accordingly intensively decreased to the end of the nozzle diverging part. Sharp decreasing of a friction out of nozzle leads again to expansion of the steam jet and one is accompanied by natural decreasing of the jet temperature and static pressure and by increasing of the jet velocity and Mach number accordingly. If superheating of steam is weak the jet motion can be accompanied by the steam condensation. In particular, co-ordinate of the thermodynamically possible beginning of the steam condensation in exhaust jet is pointed on Fig. 5.

The solution results, presented on Fig. 1, testify also that a change of the gas stream static pressure is mainly determined by a change of the stream cross-section area increased on the diameter quadrate law at the developed supersonic flow in the diverging part of Laval's nozzle. In contrast to it, a friction of the supersonic gas stream against wall of nozzle is accompanied by Joule heat and one leads to the friction thermo-kinematical effect. The essence of the effect is that the supersonic stream velocity is decreased at simultaneous increasing of its temperature and the sound velocity in it. Graphs (Fig. 1) of the stream velocity and the sound velocity in it show that increasing of length of the diverging part of the nozzle some more by 25 - 30 mm will lead the supersonic stream to the subsonic regime of motion. For all this, a lack of visible increase of static pressure is explained that increase of the nozzle diverging part cross-section considerably exceeds action of a friction onto static pressure in this case. Thus, the carried out research of Laval – Stodola nozzle profile shows that the traditional approach to calculating of the gas stream parameters and to designing of turbonozzle without taking into account of a friction is physically inferior and therefore one is traditionally doomed to expensive and long experimental working through. Considering the adduced results of computation it should be noted that the examples of reconstruction of the gas stream parameters by the results of experimental measurement of static head in the tested flow elements (systems) with the subsequent use of isentropic formulas and empirical coefficients is given in special literature on engineering gasdynamics [2, 3]. Results of the computing, presented on Fig. 1 in given article, disclose the fallacy of such traditional recommendation and ones are evidence of its superiority in comparison with industrial experiment. The computing experiment results on Fig. 2 – 6 are presented for comparison between its and with results on Fig.1.

The metering characteristics of turdonozzle with Stodola's profile and of its inlet part separately, presented on Fig. 7, 8, are different from the flow element metering characteristics offered by G. Zeuner and then W. Schule [5] on the boundary of the 19 – 20th centuries and remained to our time in educational and special literature on technical thermodynamics and gasdynamics. Difference based on principle is that Saint-Venant – Wantzel's formula in its initial kind (1839) is assumed as a base in traditional approach for determination of the flow element metering characteristics. From a point of view of physical sense the formula allow determining the gas motion velocity when the gas is expanded in the unlimited gas medium and the gas mass center remains motionless. In their previous articles [6 - 8] the authors had established the static head distribution law in the viscous compressible fluid stream along the length of flow element. Then the authors with the help of the law had led Saint-Venant – Wantzel's formula to the kind suitable for computation of the gas outflow velocity out of flow element, system. Just the methodological distinction determines a number of differences of the metering characteristics presented in given article from traditional standard. In particular, the metering characteristic of a given flow element are essentially differed in dependence on the character of the acting pressure drop. The weight flow of turbonozzle tested by Stodola grows by straight line when pressure before inlet in nozzle is increased at the supercritical pressure drops, and the weight flow is the horizontal line when the backpressure is decreased. The nozzle weight flow quantity at the subcritical pressure drops riches about 140 % in comparison with the weight flow quantity at the supercritical pressure drops. The feature is a consequence that Laval nozzle functionates in hydrodynamic regime as Venturi nozzle at the subcritical pressure drops. Laval's nozzle inlet part (without diverging part) metering characteristics is also non-linear at the subcritical pressure drops. Natural feature of the metering characteristics of Laval nozzle and other flow elements, systems is that the transition to a supercritical flow regime is clear visible on the graphs. In one of previous article [1] authors had shown that the mentioned transition is bound with the change of the gas stream motion structure.

**Final remarks**

Results adduced in the presented paper show a superiority of computing experiment in comparison with industrial experiment. Subsequent development of "The Flow System Gasdynamics" theme and VeriGas program-modeler envisages the transition from mainly one-dimensional to completely axisymmetrical statement of problem. One will allow giving quantitative evaluation of the energy 3-D distribution, the velocity profile and the velocity head profile in the viscous compressible fluid stream under action of different physical factors.

---

[1] S.L. Arsenjev, I.B. Lozovitski, Y.P. Sirik, "The Gasdynamics First Problem Solution. The Gas Stream Parameters, Structures, Metering Characteristics for Pipe, Nozzle," http://uk.arXiv.org/abs/physics/0306160 21 June 2003.

[2] Deich E. M., Engineering Gasdynamics, Energy Publishing, Moscow, 592, 1974.

[3] Stodola A., Steam and Gas Turbines, v. 1, 1927.

[4] Fundamentals of Gas Dynamics, Ed. H. W. Emmons, Princeton University Press, Princeton, New Jersey, 1958; Transl. into Russian, Foreign Literature Publishing, Moscow, 1963.

[5] Schule W., Technische warmemechanik, Verlag von J. Springer, Berlin, 1909.

[6] S.L. Arsenjev, I.B. Lozovitski, Y.P.Sirik, "The Flowing System Gasdynamics Part 1: On static head in the pipe flowing element," http://uk.arXiv.org/abs/physics/0301070, 2003.

[7] S.L. Arsenjev, I.B. Lozovitski, Y.P.Sirik, "The Flowing System Gasdynamics Part 2: Euler's momentum conservation equation solution," http://uk.arXiv.org/abs/physics/0302020, 2003.

[8] S.L. Arsenjev, I.B. Lozovitski, Y.P.Sirik, "The Flowing System Gasdynamics Part 3: Saint-Venant – Wantzel's formula modern form," http://uk.arXiv.org/abs/physics/0302028, 2003.

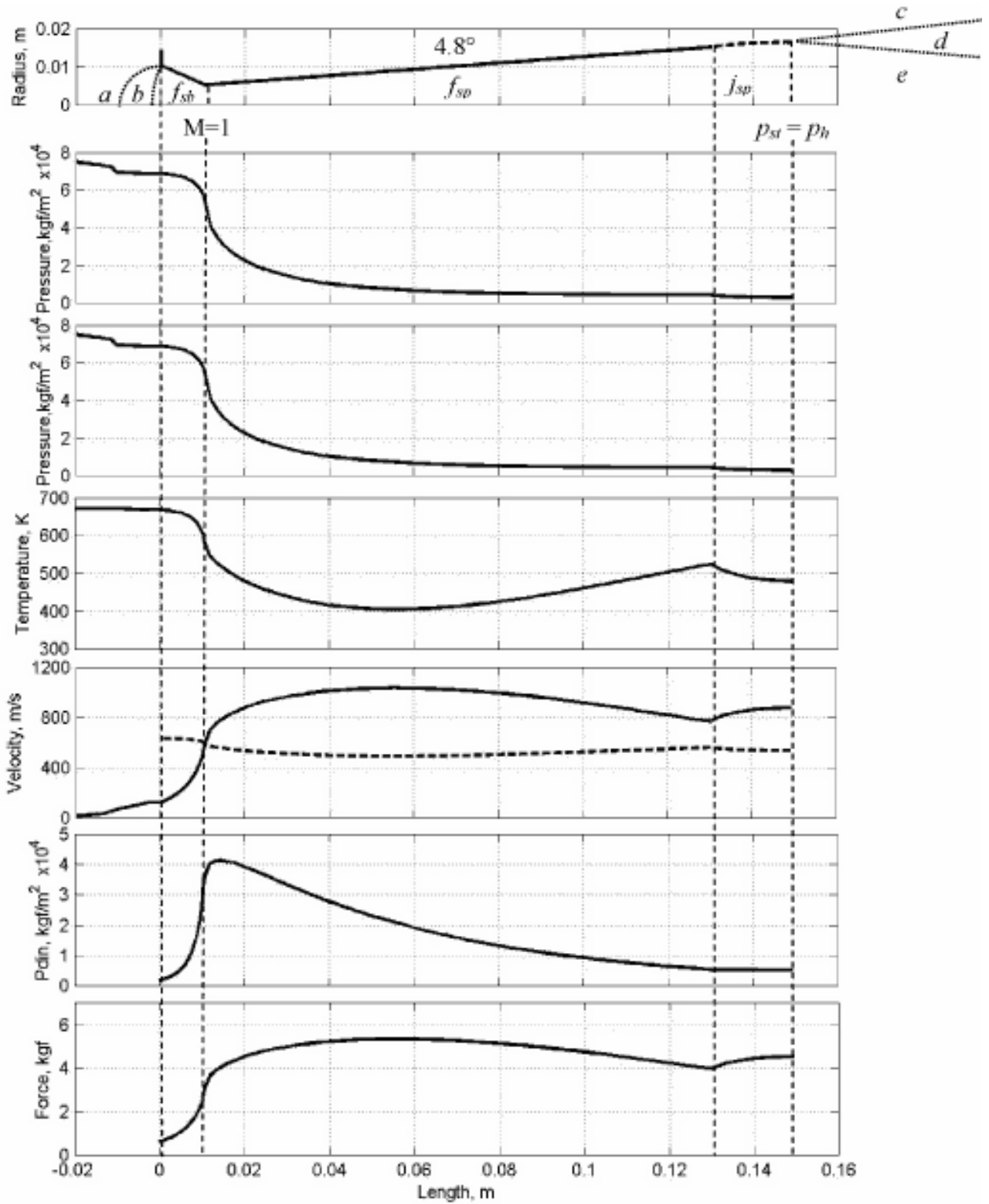

$p_0 = 10.55 \times 10^4$ kgf/m$^2$ abs const, $T_0 = 673$K    $p_h = 0.3 \times 10^4$ kgf/m$^2$

Specific impulse:  39.57    44.51 kgf·s/kg

Thermo-Supercritical, Baro-Supercritical Flow: friction (viscosity), adiabat, changing of a cross-section area

Flow mode: Laminar, Re = 0,822×10$^5$

Weight flow: 0.10109 kg/s

Fig. 1: Steam flow parameters in active nozzle (approximately Laval–Stodola profile, 1927)

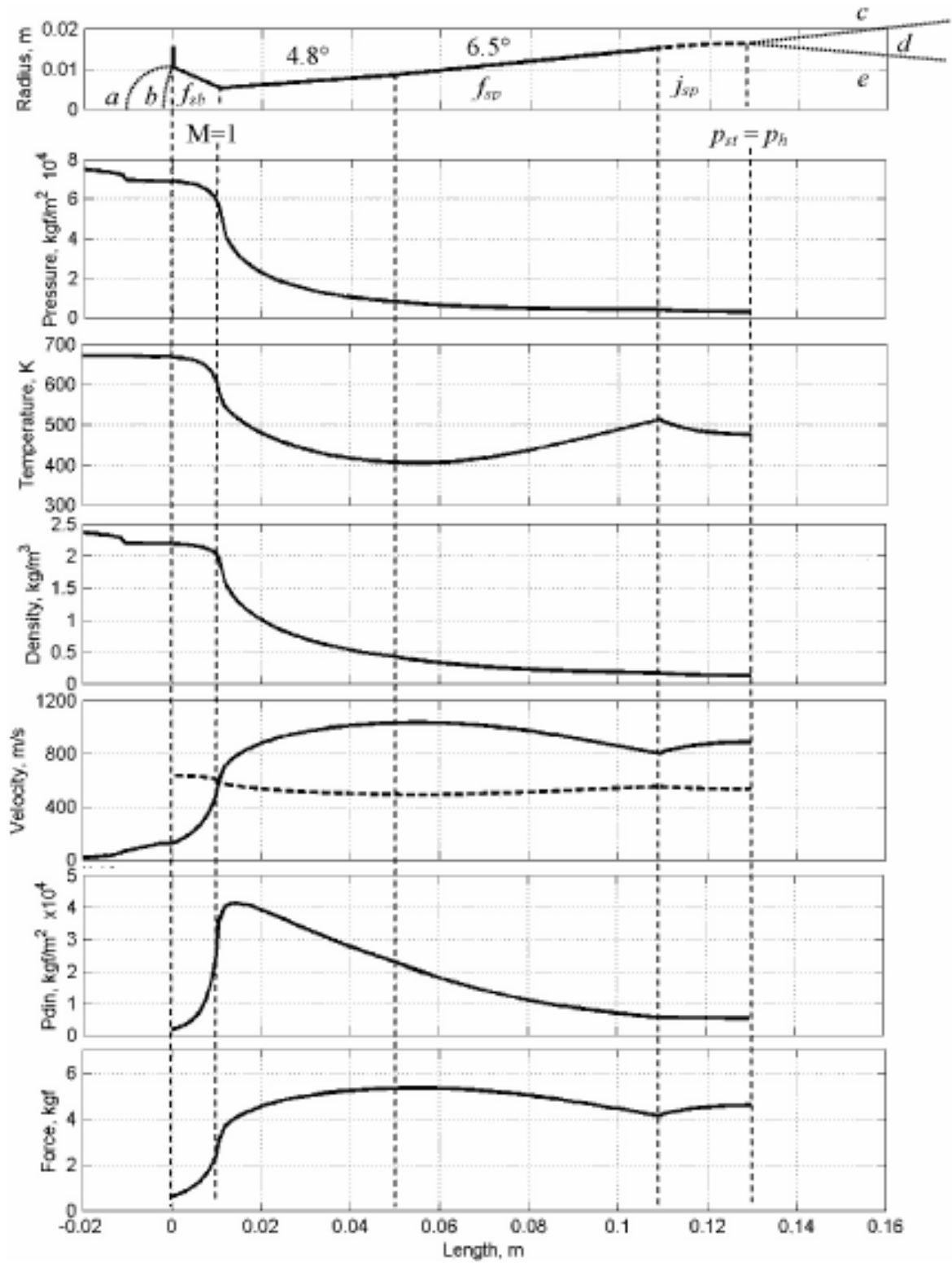

Specific impulse: 41.05   45.50 kgf·s/kg
Thermo-Supercritical, Baro-Supercritical Flow: friction (viscosity), adiabat, changing of a cross-section area
Flow mode: Laminar, Re = $0.825 \times 10^5$
Weight flow: 0.10109 kg/s

Fig. 2: Steam flow parameters in active nozzle
(biconical profile $\alpha/2 = 4.8°/6.5°$ – approximately Stanton profile)

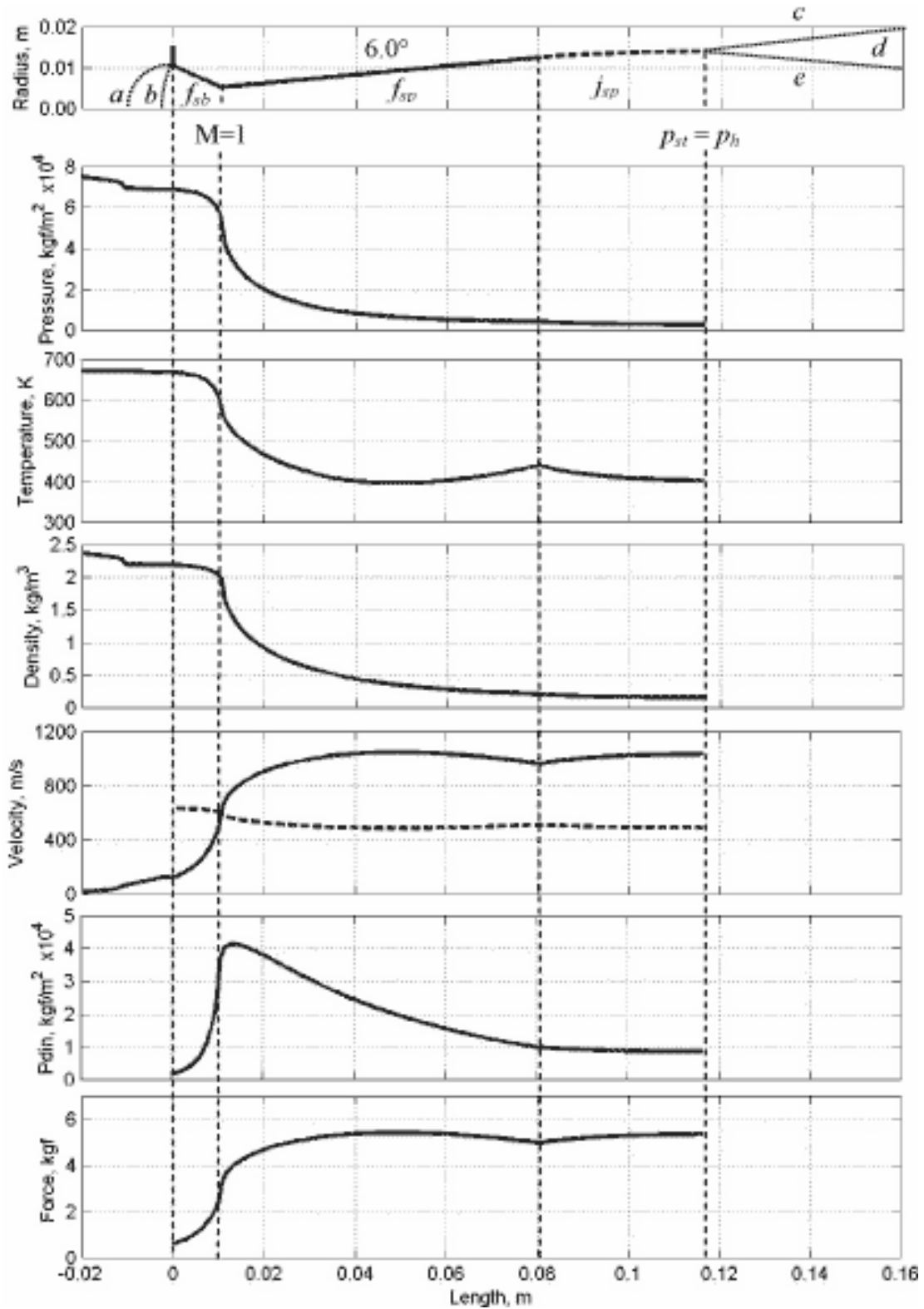

Specific impulse: 49.4    52.85 kgf·s/kg
Thermo-Supercritical, Baro-Supercritical Flow: friction (viscosity), adiabat, changing of a cross-section area
Flow mode: Laminar, Re = 0,834×10$^5$
Weight flow: 0.10109 kg/s

Fig. 3: Steam flow parameters in active nozzle
(shortened profile 1)

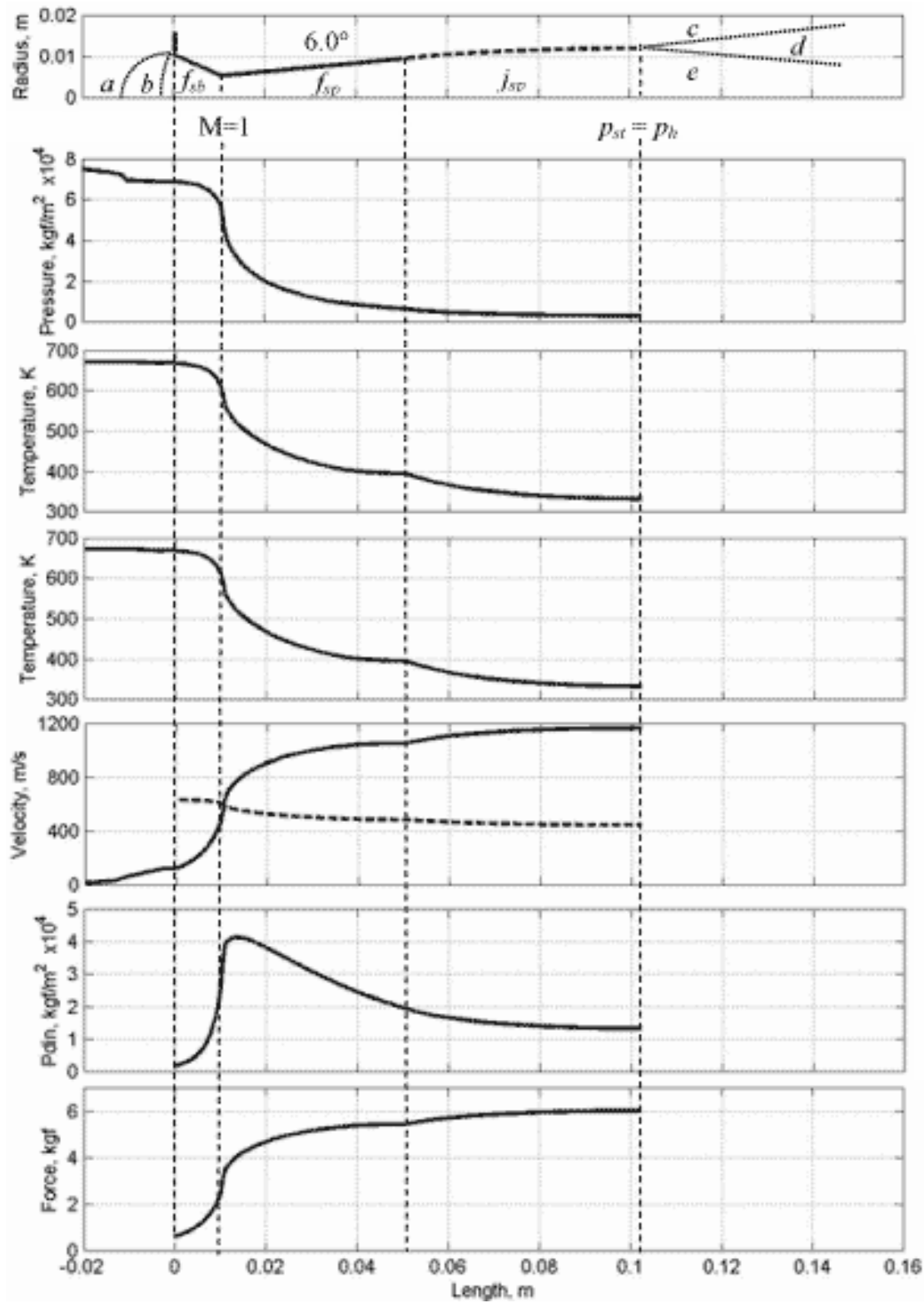

Specific impulse: 53.4   59.49 kgf·s/kg
Thermo-Supercritical, Baro-Supercritical Flow: friction (viscosity), adiabat, changing of a cross-section area
Flow mode: Laminar, Re = 0,845×10$^5$
Weight flow: 0.10109 kg/s

Fig. 4: Steam flow parameters in active nozzle
(shortened profile 2)

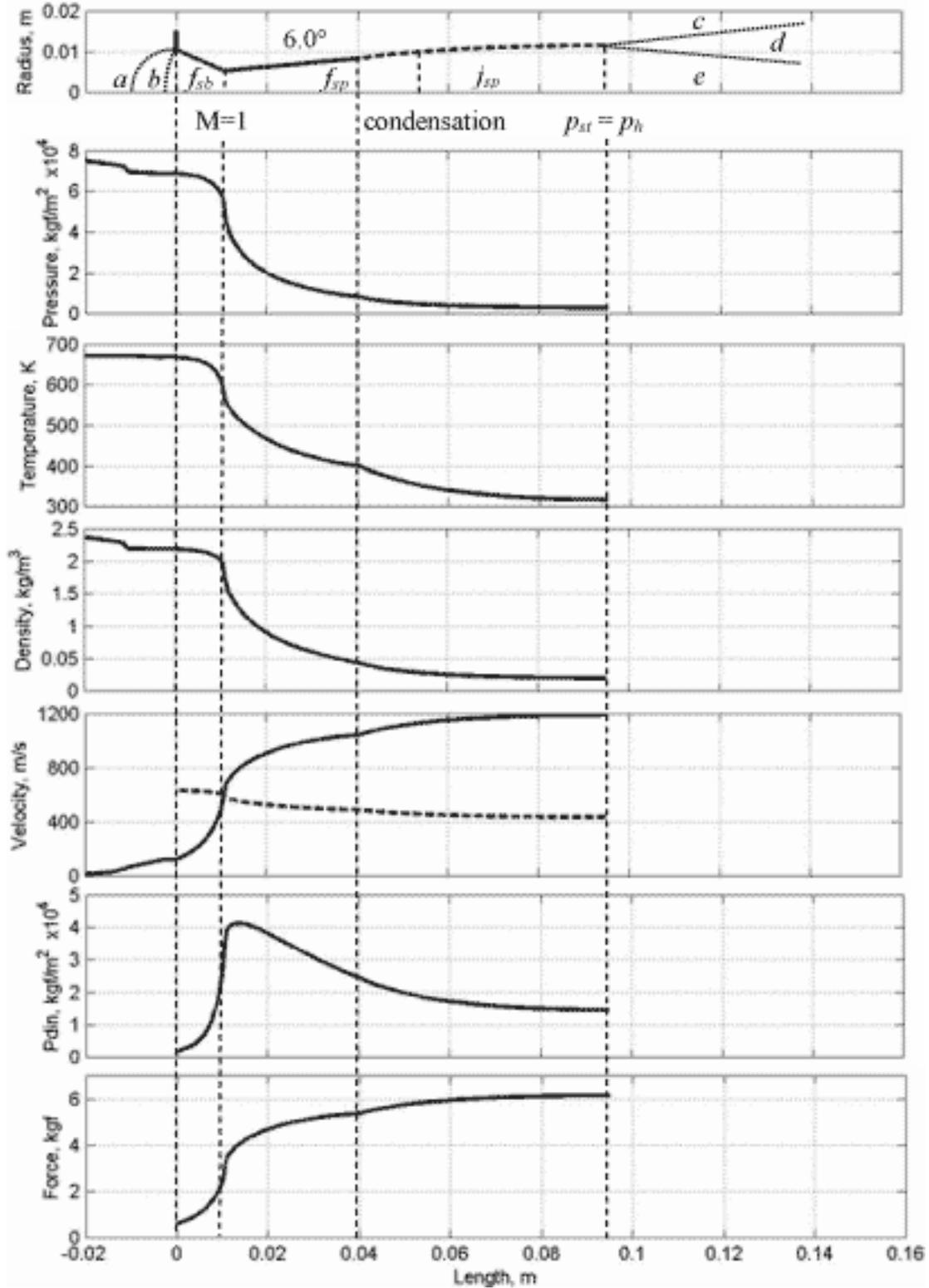

Specific impulse: 54.40    60.79 kgf·s/kg

Thermo-Supercritical, Baro-Supercritical Flow: friction (viscosity), adiabat, changing of a cross-section area

Flow mode: Laminar, Re = 0,848×10$^5$

Weight flow: 0.10109 kg/s

Fig 5: Steam flow parameters in active nozzle
(shortened profile 3)

$p_0 = 10.55\times10^4$ kgf/m² abs const, $T_0=673$K    $p_h = 0.3\times10^4$ kgf/m²

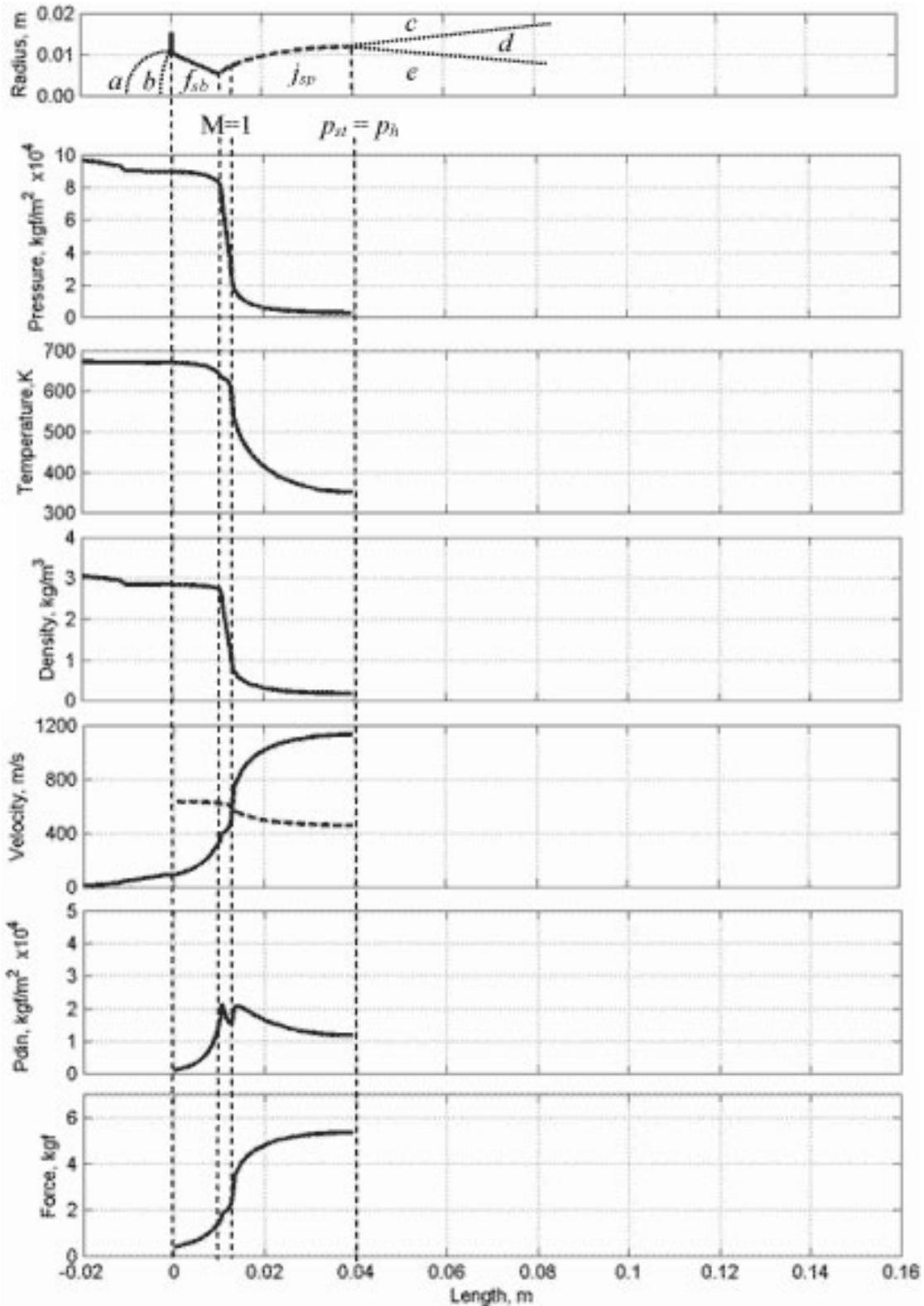

Specific impulse: 32.36    57.74 kgf·s/kg

Thermo-Supercritical, Baro-Supercritical Flow: friction (viscosity), adiabat, changing of a cross-section area

Flow mode: Laminar, Re = 0,250×10⁶

Weight flow: 0.09272 kg/s

Fig 6: Steam flow parameters in the active nozzle inlet part without diverging part

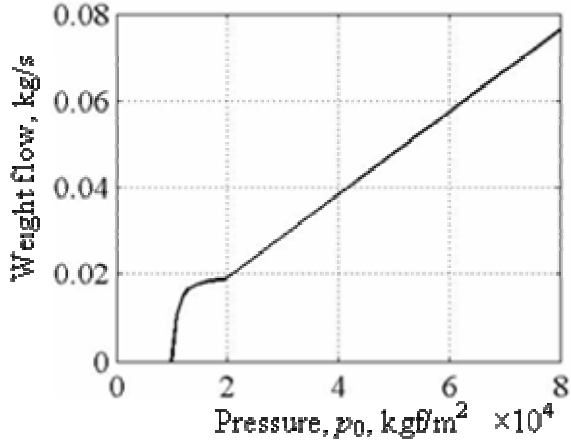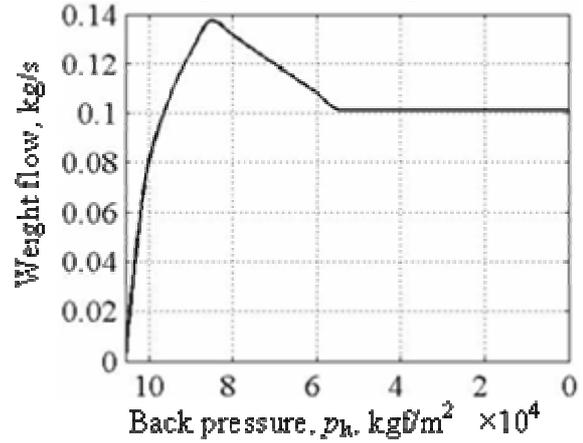

Fig. 7: Metering characteristics of active nozzle (approximately Laval-Stodola profile, 1927).
Fluid: steam $T_0 = 673$ K, adiabat, friction (viscosity), changing of a cross-section area

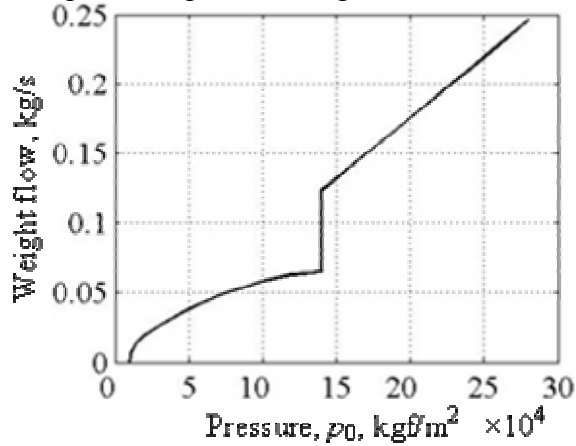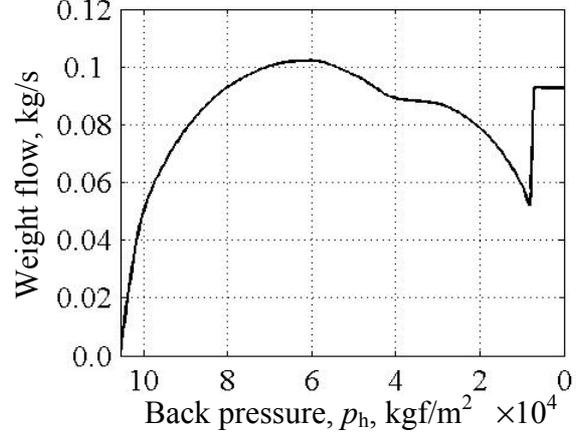

Fig. 8: Metering characteristics of the active nozzle inlet part without diverging part.
Fluid: steam $T_0 = 673$ K, adiabat, friction (viscosity), changing of a cross-section area